\documentclass{article}
\usepackage[hyphens]{url}
\usepackage{dcase2019,amsmath,graphicx,times,booktabs, tabularx}
\usepackage{hyperref}
\usepackage{float}
\usepackage{booktabs}
\usepackage{amssymb}
\usepackage{mathtools}
\usepackage[inline]{enumitem}

\usepackage{xcolor}

\title{Exploiting Parallel Audio Recordings to Enforce\\ Device Invariance in CNN-based Acoustic Scene Classification}
\name{Paul Primus$^1$, Hamid Eghbal-zadeh$^{1,2}$, David Eitelsebner$^1$ % \vspace{-2mm}
}
\secondlinename{Khaled Koutini$^{1}$, Andreas Arzt$^{1}$, Gerhard Widmer$^{1,2}$ % \vspace{-3mm}
}

\address{$^1$Institute of Computational Perception (CP-JKU) \& $^2$LIT Artificial Intelligence Lab,\\
Johannes Kepler University Linz, Austria\\
 paul.primus@jku.at 
 }
\begin{document}

\ninept
\maketitle

% \hamid{Author list needs to be extended.}

% \hamid{how about my author list and address style suggestion? the "address" is now the same as in Khaled's paper. if you need one additional line, you can skip the email address. (I also added Gerhard to the author list.)}
%\gerhard{I would drop the "End-to-end" in the title. %The title is extremely long.}

%\gerhard{The title in its current form is still rather general; it does not give a clue regarding the strategy used to obtain domain invariance.
%Examples:
%"Exploiting Parallel Audio Recordings to Obtain (better: Enforce? Encourage?) Domain %Invariance in CNN-based Audio Scene Classification"
%or:
%"Designing a Combined Loss Function to Obtain Domain %Invariance
% in CNN-based Audio Scene Classification"}

\begin{sloppy}

\begin{figure*}[t]
  \centering
    \includegraphics[width=.3\textwidth]{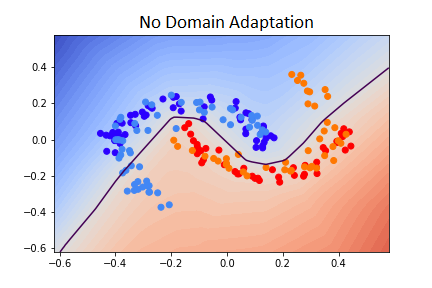}
    \includegraphics[width=.3\textwidth]{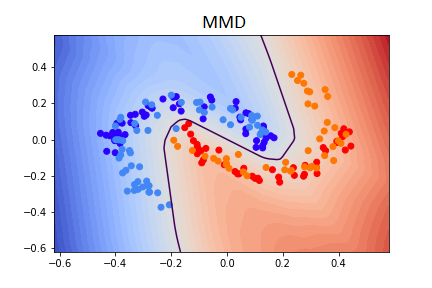}
    \includegraphics[width=.3\textwidth]{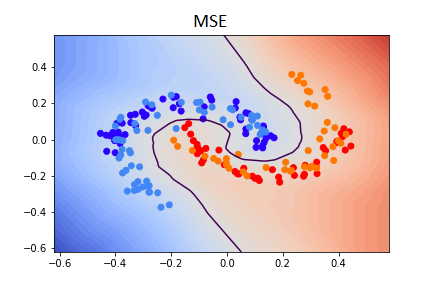}
    \caption{Two Moons dataset: best classifiers found by grid search over $\lambda$ and $n$ (Tab.~\ref{tab:two_moon}). The source domain is represented by dark blue and red data points, the shifted target domain by light blue and orange points. The black line shows the decision boundary of a classifier trained without DA (left), with MMD-DA (middle) and with MSE-DA (right). Red and blue shaded areas represent decision areas of the classifiers. }
    \label{figure:two_moons}
\end{figure*}

\begin{abstract}
% Acoustic signals carry rich information about our environment and are therefore increasingly of interest for the machine learning community.
Distribution mismatches between the data seen at training and at application time remain a major challenge in all application areas of machine learning. We study this problem in the context of machine listening (Task 1b of the DCASE 2019 Challenge). We propose a novel approach to learn domain-invariant classifiers in an end-to-end fashion by enforcing equal hidden layer representations for domain-parallel samples, i.e. time-aligned recordings from different recording devices. No classification labels are needed for our domain adaptation (DA) method, which makes the data collection process cheaper.
% \hamid{what is unlabeled here? because you have both class and device labels for training. be more specific! (or perhaps you mean testing samples?)}
We show that our method improves the target domain accuracy for both a toy dataset and an urban acoustic scenes dataset. %compared to models trained without DA.
We further compare our method to Maximum Mean Discrepancy-based DA and find it more robust to the choice of DA parameters. Our submission, based on this method, to DCASE 2019 Task 1b gave us the 4th place in the team ranking.
\end{abstract}


\section{Introduction}

\label{sec:intro}
Convolutional Neural Networks (CNNs) have become state of the art tools for audio related machine learning tasks, such as acoustic scene classification, audio tagging and sound event localization. 
 While CNNs are known to generalize well if the recording conditions for training and unseen data remain the same, the generalization of this class of models degrades when there is a distribution dissimilarity between the training and the testing data~\cite{eghbal2018deep}.
% the same does not necessarily hold if conditions change.

In the following work we elaborate our findings for subtask 1b of 2019's IEEE DCASE Challenge, % on Detection and Classification of Acoustic Scenes and Events
%\hamid{I think we need to cite the 2019 challenge report, not 2018.} 
which is concerned with a domain mismatch problem.
The task is to create an acoustic scene classification system for ten different acoustic classes. 
A set of labelled audio snippets recorded with a high-quality microphone (known as Device A) is provided for training.
Additionally, for a small subset of samples from device A, parallel recordings from two lower quality microphones (devices B and C) are given. 
Evaluation of methods is done based on the overall accuracy on unseen samples from devices B and C. 
The acoustic scene, the city, and the device labels are provided for samples of the development set only.
The main challenge of task 1b is to develop a model that, although trained mostly on samples from device A, is able to generalize well to samples from devices B and C.
Since this problem is related to the field of \emph{Domain Adaptation (DA)}, we refer to the distribution of device A samples as the \emph{source domain}, and the distribution of samples of B and C devices as the \emph{target domain}.
In this work we explain how a state-of-the-art CNN model which by itself achieves high accuracy can be further improved by using a simple DA technique designed for problems where parallel representations are given.

\section{Related Work}
Domain Adaptation (DA) is a popular field of research in transfer learning with multiple areas of application, e.g. bird audio detection~\cite{Berger_JKU}. Kouw et al.~\cite{kouw2018introduction} distinguish between three types of data shifts which lead to a domain mismatches: prior, covariate and concept shift. In this work we focus on domain mismatches which are caused by covariate shifts (i.e., changes in feature distributions).

According to Shen et al.~\cite{shen2017wasserstein} solutions to domain adaptation can be categorized into three types:
\begin{enumerate*}[label=(\roman*)]
    \item Instance-based methods: reweight or subsample the source dataset to match the target distribution more closely~\cite{huang2007correcting}.
    \item Parameter-based methods: transfer knowledge through shared or regularized parameters of source and target domain learners~\cite{rozantsev2018beyond}, or by weighted ensembling of multiple source learners~\cite{duan2012exploiting}.
    \item Feature-based methods: transform the samples such that they are invariant of the domain. Weiss et al.~\cite{weiss2016survey} further distinguish between symmetric and asymmetric methods. Asymmetric methods transform features of one domain to match another domain~\cite{hoffman2014asymmetric} symmetric feature-based methods embed samples into a common latent space where source and target feature distributions are close~\cite{tzeng2014deep}.
\end{enumerate*}
Symmetric feature-based methods can be easily incorporated into deep neural networks and therefore have been studied to a larger extent. The general idea is to minimize the divergence between source and target domain distributions for specific hidden layer representations with the help of some metric of distribution difference. For example, the deep domain confusion method~\cite{tzeng2014deep} and deep adaptaion network~\cite{long2015learning} use Maximum Mean Discrepancy (MMD)~\cite{gretton2012kernel} as a non-parametric integral probability metric.
Other symmetric feature-based approaches exist that use adversarial objectives to minimize domain differences~\cite{tzeng2017adversarial, shen2017wasserstein}. 
These methods learn domain-invariant features by playing a minimax game between the domain critic and the feature extractor where the critic's task is to discriminate between the source and the target domain samples and the feature extractor learns domain-invariant and class-discriminative features. However, training the critic introduces more complexity, and may cause additional problems such as instability and mode collapse.
% \hamid{not a big fan of this last sentence ;)}
\section{Domain-Invariant Learning}
We propose a symmetric feature-based loss function to encourage the network to learn device-invariant representations for parallel samples from the source $X^s$, and the target domain $X^t$. 
This loss exploits the fact that parallel samples contain the same information relevant for classification and differ only due to a covariate shift, e.g. time-aligned spectrograms $(x^s, x^t)$ contain the same information about the acoustic scenes and differ only due to device characteristics. Let $\phi_l(x^s)$ and $\phi_l(x^t)$ be $d$-dimensional hidden layer activations of layer $l$ for paired samples $x^s$ and $x^t$ from the source and the target domains, respectively. A domain-invariant mapping  $\phi_l(\cdot)$ projects both samples to the same activations without losing the class-discriminative power. 
To achieve this, we  propose to jointly minimize classification loss $\mathcal{L}_{\mathit{CL}}$ and the Mean Squared Error (MSE) over paired sample activations, where the latter one  is defined as
\begin{equation}
\label{eq:pmse}
\mathcal{L}_{\mathit{l, MSE}} = \frac{1}{n \cdot d}\sum_{i=1}^{n} \left\lVert \phi_l(x_i^s) - \phi_l(x_i^t)\right\rVert^2_2
\end{equation}
for some fixed network layer $l$ (this is a hyper-parameter).
As we will show in Section~\ref{sec:experiments} the DA mini-batch size $n$ is critical, and our results suggest that bigger $n$ yields better results. The final optimization objective we use for training is a combination of classification loss $\mathcal{L}_{\mathit{CL}}$ and DA loss $\mathcal{L}_{\mathit{l, MSE}}$:
\begin{equation} \label{eq:da}
\mathcal{L} = \mathcal{L}_{\mathit{CL}} + \lambda \mathcal{L}_{\mathit{l, MSE}}
\end{equation}
Here, $\lambda$ controls the balance between the DA loss and the classification loss during training. Note that for $\mathcal{L}_{\mathit{l, MSE}}$ no class label information is required and the labeled samples from all domains can be used for the supervised classification loss $\mathcal{L}_{\mathit{CL}}$.

\section{Experiments}
\label{sec:experiments}
In the following we evaluate the performance of our approach on the two moons dataset as well as on real-world acoustic data: the \emph{DCASE 2019 Task 1b dataset} on acoustic scene classification~\cite{Mesaros2018_DCASE}.
%\gerhard{You cite the 2018 DCASE report?}
We compare our proposed DA objective to the multi-kernel MMD-based approach %~\cite{long2015learning}
used by Eghbal-zadeh et al.~\cite{Eghbal-zadeh2019} for DCASE 2019 Subtask 1b. In all experiments, parallel samples are used without any class-label information. For both datasets, we find that when paired samples are given, MSE achieves higher accuracy on the target set compared to MMD.

\subsection{Experimental Setup}
We compare our approach to a baseline that uses the same CNN architecture and classification loss, but does not incorporate a DA loss. As another baseline, we use multi-kernel MMD-based DA~\cite{long2015learning}, a non-parametric symmetric feature-based approach.
%\hamid{how about: We compare our approach to a baseline that uses the same CNN architecture and classification loss, but does not incorporate a DA loss. As another baseline, we use a multi-kernel MMD-based DA~\cite{long2015learning}, a non-parametric symmetric feature-based approach.}
%The idea of MMD-based DA is to embed hidden representations of specific layers into reproducing kernel Hilbert space and minimizes the divergence between mean embeddings of source and target domain distributions\hamid{this last sentence does not read very good (embed hidden representations into reproducing kernel Hilbert space)}.
MMD represents distances between two distributions as distances between mean embeddings of features in reproducing kernel Hilbert space $\mathcal{H}_k$:
\begin{equation*}
  d_k^2(X^s, X^t) = \left\lVert \mathbb{E}_ {X^s} \big[k(\phi_l(x^s), \cdot\ )\big] - \mathbb{E}_ {X^t} \big[ k(\phi_l(x^t), \cdot\ )\big] \right\rVert_{\mathcal{H}_k}^2
\end{equation*}
The kernel $k$ associated with the feature mapping for our experiments is a combination of four equally weighted RBF kernels with $\sigma \in \{0.2, 0.5, 0.9, 1.3\}$. We use the empirical version of this metric as DA loss, for which we randomly sample batches of size $n$ from $X^s$ and $X^t$. Therefore batches do not necessarily contain parallel representations of samples. Compared to our approach, MMD-based DA matches the distribution between the hidden representations of the source and the target domains, and not between the parallel representations. For both DA methods best results were obtained when applying the DA to the output layer. A plausible explanation is that using higher layer activations gives the network more flexibility for learning domain invariant representations.

\subsection{Two Moons}
Two moons (see Fig.~\ref{figure:two_moons}) is a toy dataset often used in the context of transfer learning. It consists of two interleaved class distributions, where each is shaped like a half circle. We use this synthetic dataset to demonstrate our domain adaptation technique under controlled conditions.

\begin{table}[t]
\footnotesize
\centering
\begin{tabular}{ @{}lrrrrrrrr@{} }
\toprule
 & \multicolumn{4}{c}{MSE}  & \multicolumn{4}{c}{MMD} \\
 \cmidrule(lr{.5em}){2-5} \cmidrule(lr{.5em}){6-9} 
 $n$\,\textbackslash\,$\lambda$  & 0.1 & 1 & 5 & 10 & 0.1 & 1 & 5 & 10 \\
\midrule
8   & .999 & .999   &  .999  &  .999            & .805 & .862   & .760 &  .749 \\
32  & .999 & .999   &  .999  &  .999            & .817 & .771   & .995 &  .990 \\
128 & .999 & .999   &  .999  &  .999            & .801 & .859   & .754 &  .739 \\
256 & .999 &  .999  &  .999  &  .999            & .804 & .861   & .997 &  .744 \\
\bottomrule
\end{tabular}
\caption{Domain Adaptation (DA) results on the two moons dataset: Accuracy % (\%)
on the target domain for models trained with different choices of DA loss, $\lambda$ (columns) and $n$ (rows). Baseline without DA is at $0.814$.}
\label{tab:two_moon}
\end{table}
\begin{figure*}[ht]
  \centering
    \includegraphics[width=1\textwidth]{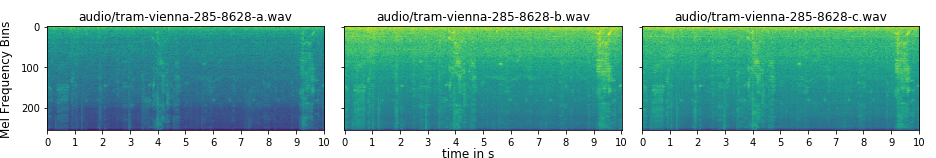}
    \caption{From left to right: Time-aligned recordings from devices A (Soundman OKM II Klassik/Studio A3 Microphone \& Zoom F8 Recorder), B (Samsung Galaxy S7) and C (iPhone SE). 
    Spectrograms show microphone-specifics, e.g. samples from devices B and C have more noise in lower Mel bins, compared to those from device A, and samples from device A seem to have fewer energy in all frequency bins.}
    \label{figure:parallel_recordings}
\end{figure*}
\subsubsection{Dataset \& Architecture \& Training}
We utilize \textit{sklearn} to generate two class-balanced two moons datasets with Gaussian distributed noise ($\mu=0$ and $\sigma=0.1$) and $10.000$ samples each. Features are normalized to fall into the range of $[-0.5, 0.5]$. 
Domain-parallel representations are obtained by applying an artificial covariate shift to one of the two datasets. 
For our initial experiments we use two transformations: a stretching along the y-dimension by a factor of 1.5, and a rotation by -45 degrees (Fig.~\ref{figure:two_moons}). We assume no label information is available for the parallel dataset. All experiments use a common model architecture, which is a fully connected network with one hidden layer of size 32 and ReLU activations. The output layer consists of one unit with a sigmoid activation function. The weights are initialized with He normal initialization~\cite{he2015delving}. We train for 250 epochs with mini-batches of size $32$, binary cross-entropy loss, ADAM~\cite{DBLP:journals/corr/KingmaB14} update rule and constant learning rate of $0.001$ to minimize Eq.~\ref{eq:da}.

\subsubsection{Results}
Without DA the model scores $81.4 \%$ accuracy on the target validation dataset (Fig.~\ref{figure:two_moons}~left). To find a good parameter setting for both domain adaptation techniques we perform grid search over the DA weight $\lambda \in \{0.1, 1, 5, 10\}$ and the DA batch size $n \in \{8, 32, 128, 256\}$. Results are summarized in Table~\ref{tab:two_moon}. At its best, MMD improves the accuracy on the target dataset to $99.7\%$ (Fig.~\ref{figure:two_moons}~middle). Regardless of the parameter combination the model with MSE-DA reaches $99.9\%$ accuracy on both source and target domain validation sets. (Fig.~\ref{figure:two_moons}~right). For all parameter configurations MSE-DA yields better results than MMD-DA.

\subsection{Urban Acoustic Scene Dataset}

The previous section has demonstrated that MSE-DA can be effectively used when domain-parallel representations are given. It is now necessary to evaluate our prior findings on a real world dataset, in our case the DCASE 2019 Task 1b dataset~\cite{Mesaros2018_DCASE}. As explained in the introduction, our objective is to create a recording device invariant classifier by training it on a larger set of source domain samples and a few
% a view to kill
time-aligned recordings from the target domain. Fig.~\ref{figure:parallel_recordings} shows three time-aligned recordings, for which we can observe the device-specific characteristics. Section~\ref{sec:intro} describes the DACSE 2019 task 1b in more details. An implementation of the following experiments is available on GitHub \footnote{\url{https://github.com/OptimusPrimus/dcase2019_task1b/tree/Workshop}}.

\label{ssec:prepare}
% almost equal to matthias's paper last year
% The distribution of samples by device and the cross validation split suggested by DCASE organizers can be seen in Table \ref{tab:dist}

\subsubsection{Dataset}
The dataset contains 12.290 non-parallel device A samples and 3.240 parallel recorded samples (1080 per device). We use the validation setup suggested by the organizers, i.e. 9185 device A, 540 device B, and 540 device C samples for training and 4185 device A, 540 device B, and 540 device C for validation. Preprocessing is done similar to~\cite{Dorfer2018}: We resample the audio signals to 22050Hz and compute a mono-channel Short Time Fourier Transform using 2048-sample windows and a hop size of 512 samples. We apply a dB conversion to the individual frequency bands of the power spectrogram and a mel-scaled filterbank for frequencies between 40 and 11025Hz, yielding 431-frame spectrograms with 256 frequency bins. The samples are normalized during training by subtracting the source training set mean and dividing by the source training set standard deviation.

%\begin{table}[t!]
%\centering
%\begin{tabular}{ @{}lrrr@{} }
%\toprule
%  Set & Device A & Device B & Device C \\
%\midrule
%Train & 9185 & 540 & 540\\
%Val  & 4185 & 540 &  540 \\
%\midrule
%Total & 13370 & 1080 &  1080 \\

%\bottomrule
%\end{tabular}
%\caption{Distribution of audio samples by devices. Training (Tain) and Validation (Val) set as split %as suggested by the challenge organizers.}
%\label{tab:dist}
%\end{table}

\label{sec:netarch}
\subsubsection{Network Architectures}
We use the model architecture introduced by Koutini et al.~\cite{Koutini2019Receptive}, a receptive-field-regularized, fully convolutional, residual network (ResNet) with five residual blocks (Tab.~\ref{tab:resnet}).
The receptive field of this architecture is tuned to achieve the best performance in audio-related tasks using spectrograms, as discussed in~\cite{Koutini2019Receptive}.

\begin{table}[ht!]
    \footnotesize
    \centering
    \begin{tabular}{ @{}lcccclc@{} }
    %\toprule
    \multicolumn{4}{c}{ResNet} & & \multicolumn{2}{c}{Residual Block (RB)}\\
    \cmidrule(lr{.5em}){1-4} \cmidrule(lr{.5em}){6-7} 
    Type & \#K & KS 1 & KS 2 & & Type & KS\\
    \cmidrule(lr{.5em}){1-4} \cmidrule(lr{.5em}){6-7} 
    Conv+BN  & 128  & 5 &   & & Conv+BN & KS 1  \\
    RB    & 128  & 3 & 1 & & Conv+BN & KS 2  \\
    Max Pool  & -    & 2 & - & & Add Input \\
    RB    & 128  & 3 & 3 & & \\
    Max Pool  & -    & 2 & - & & \\
    RB    & 128  & 3 & 3 & & \\
    RB    & 256  & 3 & 3 & & \\
    Max Pool  & -    & 2 & - & & \\
    RB    & 512  & 3 & 1 & & \\
    Conv+BN  & 10   & 3 & - & & \\
    GAP & -    & - & - & & \\
    %\bottomrule
    \end{tabular}
    \caption{Model Architecture by~\cite{Koutini2019Receptive} for experiments with the acoustic scenes dataset. \#K and KS are the number of kernels and kernel size, respectively. Residual Blocks (RB) consist of two Convolutional (Conv) layers with \#K kernels, each followed by a Batch Normalization (BN) layer. GAP is a Global Average Pooling Layer.}
    \label{tab:resnet}
\end{table}

%\begin{figure}[ht!]
%  \centering
%    \includegraphics[width=.5 \textwidth]{ResNet}
%    \caption{Model Architecture by~\cite{Koutini2019Receptive} for experiments with the acoustic scenes dataset . $k1$ and $k2$ are the kernel sizes of the first and the second layer, respectively. $d$ is the number of channels.}
%    \label{figure:resnet}
%\end{figure}

\label{ssec:Training}
\subsubsection{Training}
 Although scene labels are available for all samples, we minimize $\mathcal{L}_{CL}$ over the 8.645 non-parallel device A samples only. The 1.620 time-aligned samples are used to learn domain-invariant features by minimizing pairwise DA loss $\mathcal{L}_{l, \cdot}$ between the three devices. For each update-step, we draw a batch from the non-parallel samples and a batch from the parallel samples to compute $\mathcal{L}_{CL}$ and $\mathcal{L}_{l, \cdot}$, respectively. We then minimize the sum of these two losses (Eq.~\ref{eq:da}). Models are trained for 120 epochs with non-parallel mini-batches of size $32$, categorical cross-entropy loss, and ADAM~\cite{DBLP:journals/corr/KingmaB14} update rule to minimize Eq. \ref{eq:da}. The initial learning rate is set to $10^{-3}$ and decreased by a factor $0.5$ if the mean accuracy for devices B and C does not increase for 10 epochs. 
 If the learning rate is decreased, we also reset the model parameters to the best model in terms of mean accuracy of device B and C up to the last epoch. 
 We further use MixUp augmentation~\cite{DBLP:journals/corr/abs-1710-09412} with parameters of the beta-distribution set to $\alpha = \beta = 0.2$ for classification as well as DA samples.
 
 \subsubsection{Results}
The baseline model without domain adaptation scores  $35.3\%$ BC-accuracy. We perform grid search over parameters $\lambda \in \{0.1, 1, 10\}$ and $n \in \{1, 8, 16\}$ to find a good combination for both MMD- and MSE-DA. The best model validation accuracy on device B and C (BC-accuracy) over all 120 epochs for each experiment is reported in Table \ref{tab:asc}. MMD-DA improves the BC-accuracy compared to the baseline without DA for all except one experiment. At its best MMD-DA achieves an BC-accuracy of $49.2\%$, which is an improvement by $13.9\ p.p.$ compared to the model trained without DA. Pairwise representation matching improves BC-accuracy even further: The best MSE-DA model scores $59.2 \%$ which is $23.9 p.p.$ above the baseline without DA.
% and $23.9 p.p.$ above to the best MMD-DA model. 

\begin{table}[t]
\footnotesize
\centering
\begin{tabular}{ @{}lrrrrrrr@{} }
\toprule
 & \multicolumn{3}{c}{MSE}  & & \multicolumn{3}{c}{MMD} \\
 \cmidrule(lr{.5em}){2-4} \cmidrule(lr{.5em}){6-8} 
 $n$\,\textbackslash\,$\lambda$  & 0.1 & 1 & 10 & & 0.1 & 1 & 10 \\
\midrule
1  & .494 & .525   &  .488  &              & -     & -      & -   \\
8  & .537 & .592   &  .556  &              & .467  & .434   & .412  \\
16 & .571 & .592   &  .561  &              & .456  & .492   & .233  \\
\bottomrule
\end{tabular}
\caption{Domain Adaptation (DA) results on the acoustic scenes dataset:  Accuracy % (\%)
on devices B and C for models trained with different choices of DA loss, $\lambda$ (columns) and $n$ (rows). % on the acoustic scene dataset.
Baseline model without DA scores $.353$ accuracy on the provided split.}
\label{tab:asc}
\end{table}

% With Parallel Batches
% 1  & .494 & .525   &  .488  &              & -     & -      & -     \\
% 8  & .537 & .592   &  .556  &              & .569  & .555   & .581  \\
% 16 & .571 & .592   &  .561  &              & .488  & .560   & .560  \\

\subsection{DCASE Challenge 2019 Subtask 1b}
\label{ssec:dcase}
In the following section we describe the adjustments made to our challenge submission to be more competitive. Our technical report describes the submitted systems in more detail~\cite{Primus2019report}. 
\subsubsection{Datset \& Cross-Validation \& Training}
We split all audio segments into four folds, to have more domain parallel samples available for training. Furthermore, we minimize the classification loss over all available samples, including those from devices B and C. We increase the number of training and patience epochs to 250 and 15, respectively. For each fold, the model that scores the highest device BC-accuracy is selected for prediction on evaluation data. As we train every model on 4 folds, our final submission models are ensembles of the outputs of the 4 folds. For submission 1 and 2 we average the softmax predictions of each fold's best scoring model and select the class with the highest score. Submission 4 combines two independently trained % submission
models, again by averaging each of their 4 folds softmax outputs.

\subsubsection{Results}
\label{sec:results}
The results of our challenge submission measured in BC-accuracy on unseen samples are reported in Table~\ref{tab:dcase}. The convolutional ResNet without DA achieved a BC-accuracy of $71.3\%$ on the evaluation set, training on the suggested split achieved a BC-accuracy of $61.2\%$. 
%\hamid{Why the best results in Table 3 is .591, and in Table 4 is .64? Do they use different splits? Or only the batchsize? Maybe you mentioned, but I missed it!}
The model used in submission three trained with MSE-DA loss gained an additional $2.1 p.p.$ on the evaluation set over the base model, resulting in an accuracy of $73.4 \%$. A larger gain can be seen for the proposed split, as with $64.35$ the model performed $3.15 p.p.$ better than our base model. Our ensemble of eight predictors achieves $74.2\%$ BC-accuracy on the evaluation set which is our best result. The challenge submission by \cite{Eghbal-zadeh2019} which utilizes MMD-DA to learn device-invariant classifiers scores $74.5 \%$ on the final validation set, $0.3\ p.p.$ better than ours. The MM-DA used in~\cite{Eghbal-zadeh2019} incorporates across-device mixup augmentation, is applied on a different architecture, integrates ensemble models, and uses a larger batch size, which explains the performance differences.

\begin{table}[t]
\footnotesize
\centering
\begin{tabular}{@{}lrrrrrr@{}}
\toprule
         & Tr.\textbackslash Te.   & 4-CV  & K. Priv.      & K. Pub.       &  Eval. \\
\midrule
Ensemble &  -         & -          & \textbf{.770} & \textbf{.766} & \textbf{.742}\\
MSE-DA   &    .644    & .697       & .762          & .758          &   .734\\
No-DA    &    .612    & .669       & .705          & .737          &   .713\\
\bottomrule
\end{tabular}
\caption{DCASE 2019 Task 1b results for different validation sets, from left to right: Device B and C validation accuracy (\%) on the provided (Tr.\textbackslash Te.) and custom split (4-CV),  Kaggle private (K. Priv) and public leaderbord (K. Pub.), and the evaluation set (Eval.).}
\label{tab:dcase}
\end{table}

\section{Conclusion \& Future Work}
\label{sec:conclusion}
In this report, we have shown how an already well-performing ResNet-like model~\cite{Koutini2019Receptive} can be further improved for DCASE 2019 task 1b by using a simple DA technique. Our DA loss is designed to enforce equal hidden layer representations for different devices by exploiting time-aligned recordings. In our experiment we find that pointwise matching of representations yields better results, compared to minimizing the MMD between the hidden feature distributions without utilizing parallel representations. Notably, the MSE-DA increased the performance by 3.15 p.p. on the validation set of the proposed split, and by 2.1 p.p. on the final validation set. Furthermore, acquiring data for our method is cheap as it does not require labels for domain-parallel samples. In future work, we would like to investigate if data from unrelated acoustic scenes, i.e. scenes not relevant for classification, can be used to create device-invariant classifiers, as this would decrease cost even further.
%\hamid{something perhaps to discuss is the difference between the results here and in the challenge comparing mse and mmd. this could be argued by using different architectures (shake shake is more robust to noise) and applying mixup with DA via mmd. I randomly chose 2 devices, then from each randomly chose samples, then randomly mixed up with all other samples, then computed mmd.}
\section{Acknowledgment}
\label{sec:ack}
This work has been supported by the COMET-K2 Center of the Linz Center of Mechatronics (LCM), funded by the Austrian federal government and the Federal State of Upper Austria.
\label{sec:majhead}

% -------------------------------------------------------------------------
% Either list references using the bibliography style file IEEEtran.bst
\bibliographystyle{IEEEtran}
\bibliography{main}
%
% or list them by yourself
% \begin{thebibliography}{9}
% 
% \bibitem{dcase2016web}
%   \url{http://www.cs.tut.fi/sgn/arg/dcase2016/}.
%
% \bibitem{IEEEPDFSpec}
%   {PDF} specification for {IEEE} {X}plore$^{\textregistered}$,
%   \url{http://www.ieee.org/portal/cms_docs/pubs/confstandards/pdfs/IEEE-PDF-SpecV401.pdf}.
%
% \bibitem{PDFOpenSourceTools}
%   Creating high resolution {PDF} files for book production with 
%   open source tools, 
%   \url{http://www.grassbook.org/neteler/highres_pdf.html}.
%
% \bibitem{eWilliams1999}
% E. Williams, \emph{Fourier Acoustics: Sound Radiation and Nearfield Acoustic
%   Holography}. London, UK: Academic Press, 1999.
% 
% \bibitem{ieeecopyright}
%   \url{http://www.ieee.org/web/publications/rights/copyrightmain.html}.
%
% \bibitem{cJones2003}
% C. Jones, A. Smith, and E. Roberts, ``A sample paper in conference
%   proceedings,'' in \emph{Proc. IEEE ICASSP}, vol. II, 2003, pp. 803--806.
% 
% \bibitem{aSmith2000}
% A. Smith, C. Jones, and E. Roberts, ``A sample paper in journals,'' 
%   \emph{IEEE Trans. Signal Process.}, vol. 62, pp. 291--294, Jan. 2000.
% 
% \end{thebibliography}

\end{sloppy}
\end{document}